\documentclass[pra,showpacs,groupedaddress,amssymb,twocolumn,notitlepage,nofootinbib,longbibliography,floatfix,superscriptaddress]{revtex4-1}

\usepackage{graphicx,amsmath,dsfont}
\usepackage[dvipsnames]{xcolor}
\usepackage{mathtools}
\usepackage{float}
\usepackage[capitalize]{cleveref}

\newcommand{\ket}[1]{\left| #1 \right\rangle}
\newcommand{\bra}[1]{\left\langle #1 \right|}
\newcommand{\braket}[2]{\left\langle #1 | #2 \right\rangle}

\newcommand{\ha}{\hat{a}}
\newcommand{\had}{\hat{a}^\dagger}

\newcommand{\hbd}{\hat{b}^\dagger}

\newcommand{\op}[1]{\hat{#1}}

\newcommand{\nbar}{\bar{n}}

\newcommand{\id}{\mathds{1}}

\newcommand{\expval}[1]{\langle #1 \rangle}

\newcommand{\be}{\begin{equation}}
\newcommand{\ee}{\end{equation}}
\newcommand{\bea}{\begin{eqnarray}}
\newcommand{\eea}{\end{eqnarray}}
\newcommand{\ketbra}[2]{\left| #1 \middle\rangle \middle\langle #2 \right|}

\usepackage{color}
\usepackage{amsthm}
\usepackage{amsmath,amssymb,amsbsy}
\usepackage{xcolor}
\pagecolor{white}
\usepackage[english]{babel}
\usepackage{caption}

\usepackage{tikz}
\usetikzlibrary{arrows.meta, positioning, fit, backgrounds, calc}

\begin{document}

\title{Quantum Noise Limited Nonlinear Phase-Preserving Amplification of a Bosonic Mode}

\author{Abel te Riele}
\author{Tzula B. Propp}
\affiliation{Leiden Institute of Physics (LION), Leiden University, Niels Bohrweg 2, Leiden, 2333 CA, The Netherlands}

\date{\today}

\begin{abstract}
Does quantum mechanics allow for deterministic amplification schemes that simultaneously behave uniformly for all angles of bosonic phase space (i.e. phase preserving) and are nonlinear in the underlying field operators? Would such a scheme have some utility to quantum information science? In this paper we answer both questions with a resounding yes. Such amplifiers exist, but their form is highly constrained compared to the more general set of noise limited nonlinear amplifiers previously studied in the literature. This enables us to characterize the entire class of bosonic noise-limited nonlinear phase preserving amplifiers, and identify their utility: namely, the optimal amplification of Yurke-Stoler cat states and Kerr kitten states. 
\end{abstract}

\maketitle

\section{Quantum Amplification and Noise}

Given an ensemble of possible quantum states, what is the optimal amplification scheme? Forty years ago, Carlton Caves provided the definitive answer for equal-amplitude coherent states (linear phase preserving amplification) and for field quadrature eigenstates (linear phase sensitive amplification) \cite{Caves1982}, building both on radio engineering and microwave technology and the prior work on quantum linear amplification \cite{Johnson1928,Nyquist1928,Friis1944,Shimoda1957,Bradley1960,Haus1962,Heffner1962}. These quantum amplifiers implement the (quantum) noisy analogs of the linear optical amplifier  e.g. taking $\sin(\omega t) \to \sqrt{G}\sin(\omega t)$ with $G$ the power gain. Unlike for classical amplifiers, a quantum amplifier must preserve the bosonic commutation relation for the output field: $[\ha_{\rm out}, \had_{\rm out}] = \id$, inevitably adding at least half a quantum of noise. This fundamental noise constraint, known as the Caves limit, sets the standard against which all linear (and later, nonlinear) quantum amplifiers are benchmarked. In the same paper, Caves introduced schemes for amplification of (approximate) quadrature eigenstates (phase-sensitive amplification) and for coherent states of equal amplitude distributed on a ring in phase space (phase-preserving amplification), which have both been implemented in a variety of platforms e.g. optical \cite{Slusher1985,Levenson1985,Wu1986,Levenson1985b,Levenson1993,Ou1993}, superconducting \cite{Movshovich1990,CastellanosBeltran2007,CastellanosBeltran2008,Bergeal2010,Bergeal2010b,Hatridge2011,Macklin2015,Aumentado2020}, including extensions to probabilistic noiseless amplifiers \cite{Combes2014,Namiki2015}. The subsequent four decades of research have both deepened our understanding of linear amplification \cite{Clerk2010}, and prompted a search for nonlinear amplifiers amenable to a host of use-cases \cite{Kouznetsov1995}: photon counting \cite{Yuen1986,Yuen1986b,Ho1994,Propp2019},  Heisenberg-limited single photon detection \cite{Propp2020}, dark matter detection \cite{Kinion2011}, fermionic spin amplification \cite{Jiang2024}, and general quantum optical measurements \cite{DAriano1992,Epsteinetal2021}. 
 
The recent survey of nonlinear bosonic amplification schemes in Ref.~\cite{Epsteinetal2021}, while thorough, omits any mention of nonlinear scheme that preserves the phase of the input bosonic mode. That is, it presented (correctly at the time) that the amplification scheme implemented by two-mode squeezing resulting in the Heisenberg transformation of the annihilation operator 

\be\label{eq:CCR} 
    \ha_{\rm out} = \sqrt{G}\,\ha_{\rm in} + \sqrt{G-1}\,\hbd_{\rm in}
\ee was the only scheme known to deterministically implement a phase-preserving bosonic amplifier \cite{CavesSchumaker1985,WallsMilburn2008} . In this paper, we introduce the first schemes for nonlinear phase-preserving amplification. 

\begin{figure}[t]
    \centering
    \resizebox{\linewidth}{!}{%
    \begin{tikzpicture}[
        io/.style={font=\large},
        box/.style={draw, rounded corners=2.5pt, minimum height=1.0cm,
                    minimum width=1.2cm, align=center, font=\large},
        arr/.style={-{Latex[length=2.6mm]}, very thick},
        lbl/.style={font=\normalsize, align=center, anchor=south},
        eq/.style={font=\LARGE}
    ]
    \begin{scope}[xshift=1.65cm]
        \node[io] (ain1) at (0,0) {$\op{a}_{\rm in}$};
        \coordinate (t1L)  at (1.5,0.75);
        \coordinate (t1BL) at (1.5,-0.75);
        \coordinate (t1R)  at (3.2,0);
        \coordinate (t1mid) at ($(t1L)!0.5!(t1BL)$);
        \coordinate (t1C)  at (barycentric cs:t1L=1,t1BL=1,t1R=1);
        \draw[fill=blue!8, very thick] (t1L) -- (t1BL) -- (t1R) -- cycle;
        \node[font=\large] at (t1C) {$\sqrt{G}$};
        \node[io] (aout1) at (4.6,0) {$\op{a}_{\rm out}$};
        \draw[arr] (ain1) -- (t1mid);
        \draw[arr] (t1R) -- (aout1);
        \node[lbl] at ({(1.5+3.2)/2},1.05) {\textbf{Linear Phase-Preserving Amplifier:}};
    \end{scope}
    \begin{scope}[yshift=-3.1cm]
        \node[io] (ain2) at (0,0) {$\op{a}_{\rm in}$};
        \node[box] (U) at (1.5,0) {$\op{U}$};
        \coordinate (t2L)  at (3.2,0.75);
        \coordinate (t2BL) at (3.2,-0.75);
        \coordinate (t2R)  at (4.9,0);
        \coordinate (t2mid) at ($(t2L)!0.5!(t2BL)$);
        \coordinate (t2C)  at (barycentric cs:t2L=1,t2BL=1,t2R=1);
        \draw[fill=blue!8, very thick] (t2L) -- (t2BL) -- (t2R) -- cycle;
        \node[font=\large] at (t2C) {$\sqrt{G}$};
        \node[box] (Ud) at (6.3,0) {$\op{U}^\dagger$};
        \node[io] (aout2) at (8.0,0) {$\op{a}_{\rm out}$};
        \draw[arr] (ain2) -- (U);
        \draw[arr] (U) -- (t2mid);
        \draw[arr] (t2R) -- (Ud);
        \draw[arr] (Ud) -- (aout2);
        \node[lbl] at (3.8,1.05) {\textbf{Nonlinear Phase-Preserving Amplifier:}};
    \end{scope}
    \begin{scope}[yshift=-5.6cm, xshift=1.65cm]
        \node[eq] (eqsign) at (-1.15,0) {$=$};
        \node[io] (ain3) at (0,0) {$\op{a}_{\rm in}$};
        \coordinate (t3L)  at (1.5,0.75);
        \coordinate (t3BL) at (1.5,-0.75);
        \coordinate (t3R)  at (3.2,0);
        \coordinate (t3mid) at ($(t3L)!0.5!(t3BL)$);
        \coordinate (t3C)  at (barycentric cs:t3L=1,t3BL=1,t3R=1);
        \draw[fill=orange!25, very thick] (t3L) -- (t3BL) -- (t3R) -- cycle;
        \node[font=\large] at (t3C) {$\sqrt{G}$};
        \node[io] (aout3) at (4.6,0) {$\op{a}_{\rm out}$};
        \draw[arr] (ain3) -- (t3mid);
        \draw[arr] (t3R) -- (aout3);
    \end{scope}
    \end{tikzpicture}%
    }
    \caption{Phase-preserving amplification, either linear (top, Eq.~\eqref{eq:CCR}) or nonlinear (bottom, Eq.~\eqref{eq:PRnonlin-amp}). By sandwiching a linear amplifier (blue) between a number-operator-commuting unitary $\op U$ and its adjoint $\op U^\dagger$, we produce a nonlinear phase-preserving amplifier (orange) e.g. to amplify cat states with minimal noise.}
    \label{fig:schematic}
\end{figure}
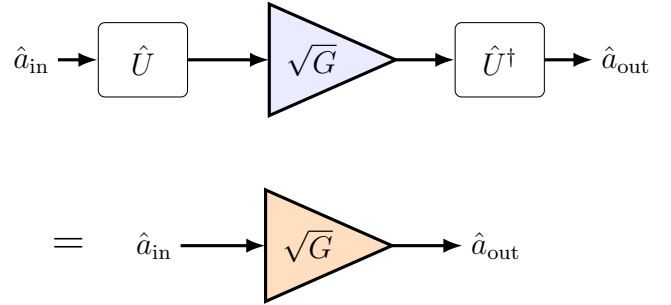

As presented in Ref.~\cite{Epsteinetal2021}, a general expression for deterministic nonlinear amplification of a single bosonic mode has the form

\be\label{eq:nonlin_amp}
    \ha_{\rm out} = \op{h}(\ha_{\rm in}, \had_{\rm in}) + \op{L}_{\rm in}
\ee with $\op{L}_{\rm in}$ a noise term necessary to preserve the commutator and $\op{h}(\ha, \had)$ an arbitrary function of the creation and annihilation operators. However, as we prove in the next section and illustrate in Fig. \ref{fig:schematic}, any deterministic nonlinear amplification of a bosonic mode \emph{must} take the form

\be\label{eq:PRnonlin-amp}
    \ha_{\rm out} = \sqrt{G}\,\op{U}\ha_{\rm in}\op{U}^\dagger + \sqrt{G-1}\,\hbd_{\rm in}. 
\ee 
Crucially, the unitary $\op{U}$ that conjugates the linear scheme in Eq. \ref{eq:CCR} must commute with the number operator: $[\op{U},\op{n}]=0$; this is prerequisite for uniform performance over all possible phases, as rotations in phase space are generated by the number operator. Operationally, this corresponds to first implementing a unitary, applying the linear scheme, and then implementing the adjoint of the first unitary as in Fig. \ref{fig:schematic}. Intuitively, this is beneficial because coherent states are robust to phase-preserving noise and thus also phase-preserving amplification. 

Our approach follows the general nonlinear amplifier form of Epstein \textit{et al.}~\cite{Epsteinetal2021}. This construction, while restrictive, exhausts the full class of deterministic bosonic single-mode nonlinear phase-preserving amplifiers and is broad enough to include schemes that optimally amplify Yurke-Stoler cat states and Kerr kitten states, as we define and illustrate in sections III and IV.

\section{The Class of Nonlinear Phase Preserving Bosonic Single Mode Amplifiers}

We now prove our first result, that the entire class of nonlinear phase preserving bosonic single mode amplifiers is generated by Eq. \ref{eq:PRnonlin-amp}. Recall from Eq.~\eqref{eq:nonlin_amp} that the most general deterministic nonlinear amplifier has the form $\ha_{\rm out} = \op{h}(\ha_{\rm in},\had_{\rm in}) + \op{L}_{\rm in}$, where $\op{L}_{\rm in}$ is a noise term arising from the internal mode of the amplifier. We model this internal mode as an independent bosonic degree of freedom whose joint state with the signal factorizes, $\op\rho = \op\rho_a\otimes\op\rho_b$, so that any operator built purely from the internal mode commutes with both $\ha_{\rm in}$ and $\had_{\rm in}$.

As in the linear case, the noise term must be independent of both $\ha$ and $\had$. This requirement, together with the condition that the commutator $[\ha, \had] = \id$ be preserved, forces the commutator of the signal function to be proportional to the identity:

\be
    [\op{h}(\ha,\had),\, \op{h}^\dagger(\ha,\had)] = c\,\id,
\ee where $c$ is real due to hermiticity.

A natural starting point is to apply unitary transformations to our Eq. \ref{eq:CCR}:

\be\label{eq:CCR_U}
    U\ha_{\rm out}U^\dagger = \sqrt{G}\,U\,\ha\,U^\dagger + \sqrt{G-1}\,U\,\hbd\,U^\dagger.
\ee

Such transformations generate nonlinear phase-preserving maps because one can apply an arbitrary unitary to define a new annihilation operator
\be\label{eq:aprime}
    \ha' = \op{U}\ha\op{U}^\dagger,
\ee which obeys the same commutation relations. To study the full class of nonlinear phase-preserving amplifiers, we ask whether there exist other maps that change the commutator while still leaving it proportional to the identity.

A natural candidate is the class of normal operators, the next larger category containing unitary operators as a subset. Such a transformation takes the form $\ha \rightarrow \op{T}\ha\op{T}^\dagger$ with $\op{T}$ a normal operator. To evaluate the commutator $[\op{T}\ha\op{T}^\dagger, \op{T}\had\op{T}^\dagger]$, we use the spectral theorem to decompose $\op{T} = \op{U}\op{D}\op{U}^\dagger$ for unitary $\op{U}$ and diagonal $\op{D}$. This reduces the problem to evaluating $[\op{D}\ha'\op{D}^\dagger, \op{D}\ha^{\prime\dagger}\op{D}^\dagger]$, where $\ha'$ is the unitarily transformed annihilation operator of \cref{eq:aprime}.

Assuming this commutator is proportional to the identity and dividing both sides by $\sqrt{c}$ (the case $c=0$ corresponds to normal operator-valued functions $\op{h}(\ha,\had)$, which are shown to be phase-sensitive in~\cite{Epsteinetal2021}), we define a new operator

\be
    \op{A} \equiv \frac{\op{D}\ha'\op{D}^\dagger}{\sqrt{c}},
    \qquad [\op{A}, \op{A}^\dagger] = \id.
\ee

Defining the Hermitian operator $\op{N} = \op{A}^\dagger\op{A}$ with eigenstates $\op{N}\ket{n} = n\ket{n}$~\cite{sakurai1994modern} and applying the commutator $[\op{N}, \op{A}] = -\op{A}$, we find

\be
    \op{N}\op{A}\ket{n} = (n-1)\op{A}\ket{n}.
\ee

Assuming no degeneracy in the spectrum, the operator $\op{A}$ thus lowers eigenvalues by exactly one, making it a valid annihilation operator on the Hilbert space. We can decompose it as

\be
    \op{A} = \op{U}'\ha'\op{U}^{\prime\dagger}
\ee for some unitary $\op{U}'$. Combining this with the definition of $\op{A}$ gives

\be
    \op{D} = c^{1/4}\op{U}',
\ee so the full normal operator $\op{T}$ reduces to a product of unitaries rescaled by the complex number $c^{1/4}$. Identifying $c = G$ recovers exactly Eq. \ref{eq:CCR}. Promoting the gain to a complex number technically introduces a constant phase shift between input and output quadratures, but this has no practical consequences.

We thus conclude that the class of nonlinear phase-preserving amplifiers (with non-normal $\op{h}(\ha,\had)$) is fully described by unitary transformations of Eq. \ref{eq:CCR} with complex gain:

\be\label{eq:nonlin_CCR}
    \ha_{\rm out} = \sqrt{G}\,\op{U}\ha_{\rm in}\op{U}^\dagger + \sqrt{G-1}\,\hbd_{\rm in}.
\ee

\section{What Makes an Amplification Scheme Optimal?}

Optimality is, at its core, subjective; what you want from an amplification scheme depends on what you want from the output signal or state. For amplifying cat and kitten states, we take this to mean preserving Wigner negativity, since it is precisely the interference structure responsible for this negativity that carries the non-classically interesting---and computationally useful---character of these states. Two quantities let us quantify this. The purity $\mu = \mathrm{Tr}(\op\rho^2)$ is connected directly to the Wigner function via $\mu = \pi\iint W^2(x,p)\,dx\,dp$, so that for fixed mean photon number, a higher purity means better preserved interference fringes and therefore greater Wigner negativity for a given cat state amplitude. The Mandel $Q$ parameter, meanwhile, is left invariant by any unitary function of $\op n$ -- in particular by the Kerr unitaries used throughout this paper -- and for a displaced thermal state of fixed amplitude is set entirely by the thermal occupation $\nbar$ of the bosonic mode; thus, a lower $Q$ corresponds directly to less added noise in the sense of Caves. Purity and $\nbar$ (via $Q$) therefore provide exactly the two handles needed to transfer the Caves bound on linear amplification into optimality conditions for the nonlinear case.

Now that we have reduced the class of nonlinear phase preserving amplifications to the form given in Eq. \eqref{eq:nonlin_CCR}, we can examine which specific amplification schemes within the class are optimal for amplifying the states of interest in this paper. In this section we show two conditions for optimal amplification of superpositions of coherent states. First, the purity of the resulting state must saturate the upper bound on the purity of a linearly amplified coherent state. Second, the Mandel $Q$ parameter \cite{Mandel1979} must be equal to that of a displaced thermal state with $\nbar = G - 1$.  The first condition can be used to show that for fixed $\nbar$, the Wigner negativity is maximal.
Optimal preservation of Wigner negativity can then be proven by showing that the Wigner negativity is monotonically decreasing in the Mandel $Q$ parameter, and by extension in $\nbar$.

\begin{figure*}[t]
    \centering
    \includegraphics[width=\textwidth]{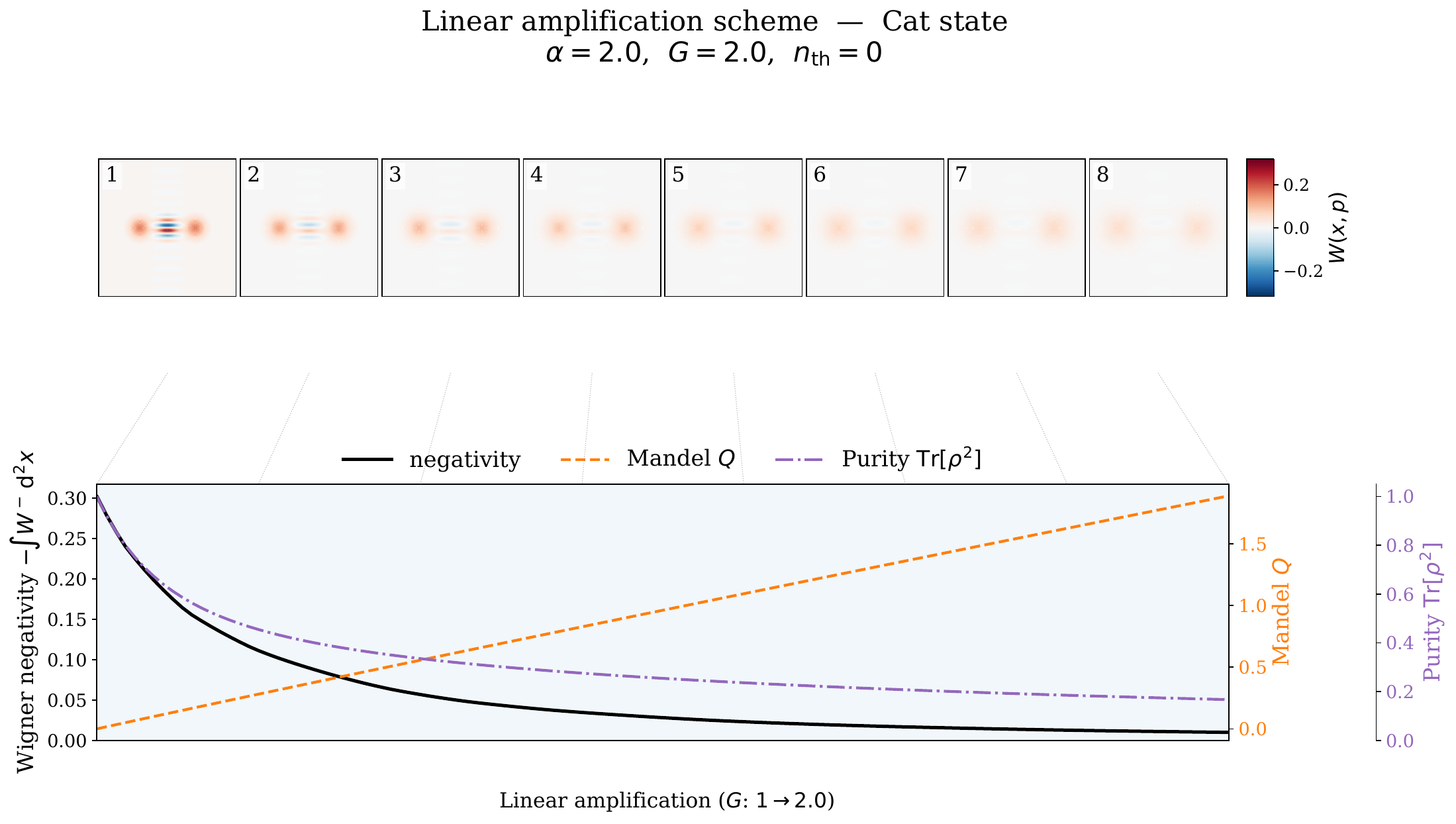}
    \caption{Evolution of the cat state $\mathcal{N}(|\alpha\rangle + i|-\alpha\rangle)$
under linear phase-preserving amplification ($\alpha=2.0$, $G\in[1,2]$,
$n_\mathrm{th}=0$). Wigner functions at eight equally spaced values of $G$
(top) and the corresponding Wigner negativity, Mandel $Q$, and purity
$\mathrm{Tr}[\rho^2]$ (bottom). Linear amplification monotonically destroys
the quantum coherence of the cat state, leaving a near-classical mixed output.}
    \label{fig:fig_timeslices_lin}
\end{figure*}

\textit{Mandel $Q$ parameter}
First, we derive a lower bound on the Mandel $Q$ parameter for amplification of states that are unitarily generated from coherent states, such as Yurke-Stoler cat states and Kerr kitten states.

To see this, we first recognize that the transformation in Eq.\eqref{eq:nonlin_CCR} can also be written as a product of two operators:

\begin{equation}
    (\op{U} \otimes \op{I}_b)\op{S}\op{a}_{in}\op{S}^\dagger(\op{U}^\dagger\otimes\op{I}_b),
\end{equation} where $\op{S}$ is the two mode squeezing operator, which means $\op{S}^\dagger\op{a}\op{S}$ gives exactly Eq.\eqref{eq:CCR}. $\op{U}$ is the general unitary function of the number operator and $\op{I}_b$ is the identity operator acting on the internal mode. Next, we recognize that the two mode squeezing operator acting on a coherent state results in a displaced thermal state. To see this explicitly, see the appendix. This motivates putting the amplification of states unitarily generated from coherent states in the following form:

\begin{multline}
    A:\{\op{U} \ket{\alpha_i} \bra{\alpha_i} \op{U}^\dagger \}\mapsto \\
    \{\op{U}\op{D}(\sqrt{G}\alpha_i)\op{\rho}_{th}(\nbar')\op{D}^\dagger(\sqrt{G}\alpha_i)\op{U}^\dagger\}.
\end{multline}

Having said this, we can continue to the proof of the lower bound on the Mandel $Q$ parameter:

\begin{equation}\label{eq:Q_opt}
    Q_{opt} = \frac{\nbar\left(\nbar + 2|\beta|^2\right)}{\nbar + |\beta|^2},
\end{equation}where $|\beta|^2 = |\alpha|^2 G$ is the squared amplitude of the displaced output field and $\nbar = (G - 1)(n_{th} + 1)$. This is precisely the Mandel $Q$ parameter of a displaced thermal state.

\textit{Proof}
Consider a map:
\begin{multline}\label{eq:amp_A}
    A:\{\op{U}\ketbra{\alpha_i}{\alpha_i}\op{U}^\dagger\}
    \mapsto \\
    \{\op{U}\op{D}(\sqrt{G}\alpha_i)\op{\rho}_{th}(\nbar')
    \op{D}^\dagger(\sqrt{G}\alpha_i)\op{U}^\dagger\}.
\end{multline}
Note that an ideal (noiseless) amplifier has $\nbar' = 0$, corresponding to a pure displaced state. Now suppose this map beats the bound, that is $Q_A < Q_{opt}$. 

We now conjugate both sides of the amplifier Eq.~\eqref{eq:amp_A} by $\op{U}$. Since $\op{U}$ is a unitary function of the number operator, it commutes with $\op{n}$ and therefore leaves the Mandel $Q$ parameter unchanged. This yields the map:
\begin{equation}
    A':\{\ketbra{\alpha_i}{\alpha_i}\}
    \mapsto
    \{\op{D}(\sqrt{G}\alpha_i)\op{\rho}_{th}(\nbar')
    \op{D}^\dagger(\sqrt{G}\alpha_i)\},
\end{equation} which is exactly the linear phase-preserving amplification of Eq.~\eqref{eq:CCR} and Ref.~\cite{Caves1982}. Since this is a displaced thermal state, the inequality becomes:
\begin{equation}\label{eq:Q_ineq}
    \frac{\nbar'\left(\nbar' + 2|\beta|^2\right)}{\nbar' 
    + |\beta|^2} < \frac{\nbar\left(\nbar + 
    2|\beta|^2\right)}{\nbar + |\beta|^2}.
\end{equation}
Since the function $f(x) = x(x + 2|\beta|^2)/(x + |\beta|^2)$ is strictly increasing in $x$ for $x \geq 0$, this implies $\nbar' < \nbar = (G -1 )(n_{th} + 1)$. The Caves limit requires $\nbar  \geq (G - 1)(n_{th} + 1)$ for any phase-preserving linear amplifier. Therefore the map $A'$ with $Q_{A'} = Q_A < Q_{opt}$ cannot exist, and neither can the map $A$.

We conclude that $Q_{opt}$ is a lower bound for phase-preserving amplification of any unitarily transformed coherent state. Furthermore, any scheme $A^*$ that 
saturates this bound is optimal for preserving any quantity that is monotonically decreasing in $Q$. $\square$

\textit{Purity}
Similar to the lower bound on the Mandel $Q$ parameter, we can derive an upper bound on the purity of the amplified state. Using the same general form for amplification as given in Eq. \ref{eq:amp_A}, we can see that the maximal allowed purity for this amplification is the same as that of linear phase preserving amplification, since purity is not affected by any unitary operation. Similar to the above proof we can conjugate both sides of the amplifier to obtain linear phase preserving amplification, without changing the purity. This shows that any phase preserving amplification must have a purity of at most $\mu = \frac{1}{2(G-1)(n_{th} + 1)}$. In Appendix \ref{app:purity} we explicitly calculate the purity of an amplified state from the Wigner function.

Figure~\ref{fig:fig_timeslices_lin} illustrates the consequence of applying this linear phase-preserving amplification directly to a Yurke-Stoler cat state. As $G$ increases, the Wigner negativity decays to zero while the Mandel $Q$ parameter grows and the purity falls, tracking the same displaced-thermal-state behavior derived above. The fine interference structure responsible for the cat state's non-classicality is washed out well before the state is appreciably amplified, motivating the nonlinear scheme we introduce next.

\section{Optimal Amplification of Yurke-Stoler Cats and Kerr kittens}

The Kerr interaction arises from a third-order nonlinear susceptibility $\chi^{(3)}$, resulting in a photon-number-dependent phase shift with Hamiltonian $\op{H}_{\rm Kerr} = \hbar\omega\op{n} + \hbar\frac{\kappa}{2}\op{n}^2$. Working in the rotating frame and absorbing the residual linear phase, the effective Hamiltonian is $\op{H}_{\rm Kerr} = \hbar\frac{\kappa }{2}\op{n}(\op{n}-1)$, generating the unitary
\begin{equation}\label{eq:kerr_unitary}
    \op{U}_{\rm Kerr}(\kappa t) = \exp\!\left[-i\frac{\kappa t}{2}\,\op{n}(\op{n}-1)\right].
\end{equation}

\begin{figure*}[t]
    \centering
    \includegraphics[width=\textwidth]{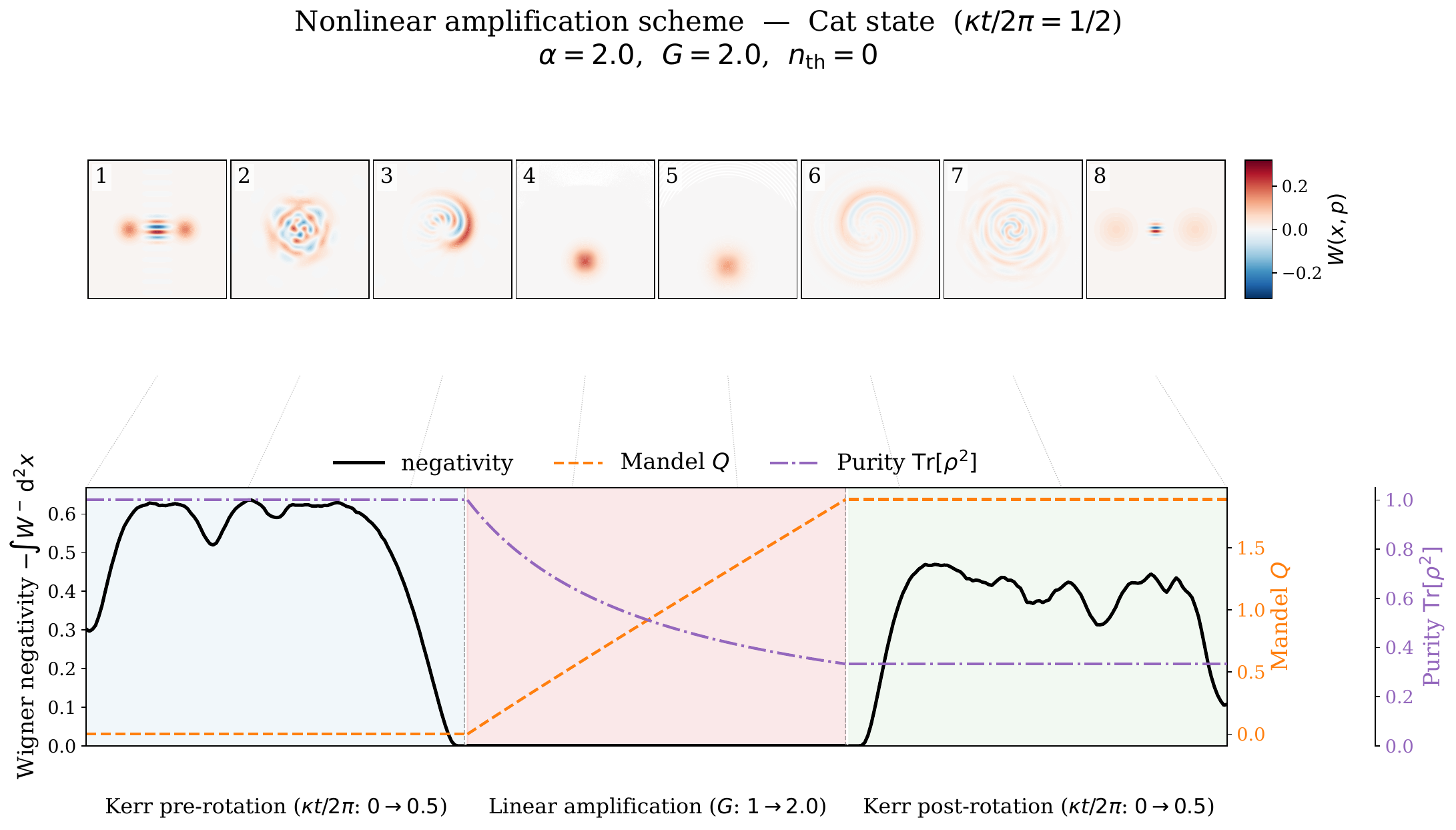}
    \caption{Evolution of the cat state $\mathcal{N}(|\alpha\rangle + i|-\alpha\rangle)$
through the nonlinear scheme $\mathcal{A}=U_K(\pi)\,\mathcal{L}_G\,U_K(\pi)$ ($\alpha=2.0$, $G=2$, $n_\mathrm{th}=0$). Wigner functions at eight equally spaced moments across the three stages (top) and the corresponding Wigner negativity, Mandel $Q$, and purity $\mathrm{Tr}[\rho^2]$ (bottom), with background shading distinguishing the Kerr pre-rotation, linear amplification, and Kerr post-rotation stages. By sandwiching the linear amplifier between two Kerr rotations, the scheme preserves significantly more negativity and purity than the linear scheme alone.}
    \label{fig:fig_timeslices_nl}
\end{figure*}

As shown by Yurke and Stoler, Eq.~\eqref{eq:kerr_unitary} evolves an input coherent state into a two-component cat state at $\kappa t = \pi$ \cite{YurkeStoler1986}; more generally, at rational fractions $\kappa t = 2\pi M/N$ it produces an $N$-component kitten state \cite{Tara1993,Tanas2003}. Since $\op{U}_{\rm Kerr}$ is diagonal in the Fock basis, it commutes with the number operator and hence, as established in the previous section, is precisely the kind of unitary that generates a nonlinear phase-preserving amplifier via Eq.~\eqref{eq:nonlin_CCR}.

These cat and kitten states owe their non-classicality to fine, sub-Planck-scale interference structure in phase space \cite{Zurek2001}, and it is exactly this structure that decoheres first, and increasingly quickly, as the superposition becomes more macroscopic \cite{Zurek1991,Propp2023}. Wigner negativity is therefore not only our figure of merit for optimal amplification, but also the quantity decoherence is most efficient at destroying -- making the preservation problem addressed here directly relevant to cat-qubit architectures built from the Yurke-Stoler family \cite{Mirrahimi2014,Grimm2020}, where the same negativity underlies any computational advantage \cite{MariEisert2012}. Though our work quantifies amplification optimality via Wigner negativity, cat states are also capable of producing entanglement when incident on a beam splitter \cite{vanEnk2003}; here, improved amplification of the cat also improves the entanglement present in the output state (and thus also increases its sensitivity to decoherence \cite{vanEnk2005}).

We now present an amplification scheme that, in comparison to the existing general method \cite{Joo2016}, is optimal for preserving the Wigner negativity of superpositions of coherent states that are unitarily generated from coherent states, i.e. Yurke-Stoler cat states and Kerr kittens:

\begin{equation}\label{eq:kitte}
    \ket{\psi_{M,N}} = \sum_{k=0}^{2N-1} f_k \ket{\alpha e^{ik\pi/N}}
\end{equation} with $N$ the number of lobes (kittens) in the superposition.

The scheme consists of two steps:
\begin{enumerate}
    \item Apply the nonlinear amplification scheme \ref{eq:nonlin_CCR} with $\op{U} = \op{U}_{\rm Kerr}(-2\pi M/N)$, mapping $\ket{\psi_{M,N}}$ to a coherent state.
    \item Apply a second Kerr unitary with opposite angle $\op{U}_{\rm Kerr}(2\pi M/N)$ i.e. the adjoint of the first unitary.
\end{enumerate}

\textit{Saturation of the bounds} In the first step, the Kerr unitary acts on the kitten state. For $\kappa t = -2\pi M/N $, it maps the kitten state to a coherent state. The Mandel $Q$ parameter remains $Q = 0$ at this point. Linear amplification then maps this to a displaced thermal state with $\nbar = G-1$, in accordance with the Caves limit, yielding a Mandel $Q$ exactly equal to $Q_{ opt}$. The second Kerr unitary maps the state to a noisy kitten-like density matrix while leaving $Q$ unchanged. The lower bound is therefore saturated. Furthermore, both Kerr unitaries leave the purity unchanged. The purity is therefore equal to that of a linear phase preserving amplification of a coherent state, thus saturating the upper bound.$\square$

To see exactly how this works, we apply the amplification scheme to the Yurke-Stoler cat state:
\begin{equation}
    \ket{\psi_{cat}} = \frac{1-i}{2}\ket{i\alpha} + \frac{1+i}{2}\ket{-i\alpha}.
\end{equation}

\textit{Stage 1: Kerr pre-rotation.}
The Kerr unitary acts on each coherent state component via the cat-splitting identity derived in appendix \ref{app:kerr-splitting}:
\begin{equation}\label{eq:catsplit}
    \op{U}_K\ket{\beta} = \frac{e^{-i\pi/4}}{\sqrt{2}}\ket{i\beta} 
    + \frac{e^{i\pi/4}}{\sqrt{2}}\ket{-i\beta}.
\end{equation}
Applying Eq.~\eqref{eq:catsplit} to each component of 
$\ket{\psi_{cat}}$ and collecting terms, the $\ket{\beta}$ 
components cancel exactly and one obtains the pure coherent 
state $\ketbra{\beta}{\beta}$, using $\mathcal{N}^2 = 1/2$. This stage can be seen in panel 1-3 of Fig. \ref{fig:fig_timeslices_nl}

\textit{Stage 2: Linear amplification.}
Linear phase-preserving amplification maps the coherent 
state to a displaced thermal state with $\nbar = G - 1$, 
saturating the Caves limit~\cite{Caves1982}(panels 4--5 of Fig.~\ref{fig:fig_timeslices_nl}). Note that here we take the internal mode to start in vacuum ($n_{th} = 0$). Using the 
Glauber--Sudarshan $P$-representation \cite{Glauber1963}
\cite{Sudarshan1963} of the thermal state, the post-linear-amplification 
density matrix is the Gaussian mixture:
\begin{equation}\label{eq:Prep}
    \op{\rho}_{dth} = \frac{1}{\pi\nbar}\int d^2\eta\; 
    e^{-|\eta|^2/\nbar}
    \ketbra{\eta - \sqrt{G}i\alpha}{\eta - \sqrt{G}i\alpha}.
\end{equation}

\textit{Stage 3: Kerr post-rotation.}
Applying Eq.~\eqref{eq:catsplit} to each coherent state 
in Eq.~\eqref{eq:Prep} with $\beta = \eta - \sqrt{G}i\alpha$ 
gives a density matrix with four terms per coherent state 
in the mixture:
\begin{multline}\label{eq:rhoK}
    \op{\rho}_{out} = \frac{1}{\pi\nbar}\int d^2\eta\;
    \frac{e^{-|\eta|^2/\nbar}}{2} \\
    \times\Big(
    \ketbra{i\beta}{i\beta}
    + \ketbra{-i\beta}{-i\beta}
    - i\ketbra{i\beta}{-i\beta}
    + i\ketbra{-i\beta}{i\beta}
    \Big).
\end{multline}
In phase space, the Wigner function of this final state looks like a cat state, but with Gaussian lobes that are displaced and broadened. This final stage corresponds to panels 6--8 of Fig.~\ref{fig:fig_timeslices_nl}, where the Wigner negativity recovers from its near-zero value at the end of the linear-amplification stage, in contrast to the monotonic decay seen for linear amplification in Fig.~\ref{fig:fig_timeslices_lin}; in comparing Fig.~\ref{fig:fig_timeslices_nl} and Fig.~\ref{fig:fig_timeslices_lin}, we see our nonlinear amplification scheme results in a higher purity and Wigner negativity given the same final Mandel Q parameter and amplification factor $G$.

\section{Conclusions}

In this work, we have resolved two open scientific questions. First, we have characterized the full class of deterministic
nonlinear phase-preserving single-mode bosonic amplifiers, proving them to be equivalent to unitary transformations of the linear phase-preserving amplifier (Eq. \ref{eq:CCR}), closing a gap in the categorization of quantum amplifiers. Secondly, within this class we have identified the optimal deterministic scheme for amplifying Yurke-Stoler cat states, proved its optimality rigorously as quantified by Wigner negativity, and suggested the same scheme be used to amplify Kerr kitten states as well. The scheme is as follows (Fig. Fig.~\ref{fig:fig_timeslices_nl}): sandwich linear amplification between two Kerr unitaries such that the linear amplification acts on a coherent state, and the quantum coherence of the cat (kitten) state is preserved as well as the laws of physics allow. Such a procedure could be helpful for situations where cat states are used to store quantum information i.e. so-called cat qubits \cite{MariEisert2012,Mirrahimi2014,Grimm2020}, provided they are chosen from the Yurke-Stoler cat family, or in quantum sensing experiments using cat states for enhanced sensitivity \cite{Zheng2025}.

Our work opens the door to several future explorations of non-linear phase-preserving amplifiers. Firstly, probabilistic non-linear amplifiers that preserve phase in complete analogy with the well-studied probabilistic noiseless linear amplifiers \cite{Combes2014,Namiki2015}. Secondly, choosing the unitary $\op{U}$ differently e.g. by using higher-order transformations beyond the Kerr interaction to generate phase-preserving amplification schemes amenable to different quantum states. One can also choose $\op{U}$ to commute not with the number operator $\op{n}$, but with generators of other symmetries of interest for a particular ensemble of quantum states, forming a large subclass of the amplifiers identified in Ref.~\cite{Epsteinetal2021}. Lastly, the question of fermionic phase preserving amplification is left completely open in this work, and we hope our work will be generalized beyond bosons by future researchers interested in probing the fundamental limits of quantum amplification.

\section{Acknowledgements}

We are grateful to Wolfgang L\"{o}ffler and Steven van Enk for helpful conversations about this project. TzBP gratefully acknowledges support from the Quantum Software Consortium Ada Lovelace Fellowship. 
\bibliography{references}

\appendix

\section{Linear Amplification of a Coherent State}\label{app:Linear Amplification of a Coherent State}
If we wish to evolve the density matrix of a coherent state, we must use the unitary transformation that is equal to the Bogoliubov transformation in Eq.~\eqref{eq:CCR}. The unitary that does this is called the two-mode squeezing operator \cite{CavesSchumaker1985}:

\begin{equation}
    \op{S} = \exp\left(r\left(\op{a}^\dagger\op{b}^\dagger 
    - \op{a}\op{b}\right)\right).
\end{equation}

This operator is where the signal mode ($\op{a}$) and the internal mode ($\op{b}$) interact. After we calculate the effects of this interaction we can trace out the internal mode to see the effect on just the signal mode. 

First, we use the fact that a coherent state can be written as a displaced vacuum state.
\begin{equation}
    \ket{\alpha} = \op{D}(\alpha)\ket{0},
\end{equation}
where the displacement operator is defined as:
\begin{equation}
    \op{D}(\alpha) = \exp\left(\alpha\op{a}^\dagger 
    - \alpha^*\op{a}\right).
\end{equation}

For an internal mode that starts in vacuum, the full input density matrix becomes:

\begin{equation}
    \rho_{in} = \op{D}_a(\alpha)(\ket{0}_a\bra{0}_a \otimes \ket{0}_b\bra{0}_b) \op{D}_a^\dagger(\alpha),
\end{equation}
where we have rearranged the displacement operator. This is valid, since it only acts on the signal mode.

We now evolve the full density matrix. Applying the two-mode squeezing operator:
\begin{equation}
    \op{\rho}_{out,total} = \op{S}\op{\rho}_{in}\op{S}^\dagger
    = \op{S}\op{D}_a(\alpha)
    \left(\ket{0,0}\bra{0,0}\right)
    \op{D}_a^\dagger(\alpha)\op{S}^\dagger.
\end{equation}
We focus on the action of $\op{S}$ on the ket side. Inserting $\op{S}\op{S}^\dagger = \op{1}$ between $\op{S}$ and $\op{D}_a(\alpha)$:
\begin{equation}
    \op{S}\op{D}_a(\alpha)\ket{0,0} = 
    \left(\op{S}\op{D}_a(\alpha)\op{S}^\dagger\right)
    \op{S}\ket{0,0}.
\end{equation}
We evaluate the conjugated displacement operator using the transformation \eqref{eq:CCR} derived above. Since 
$\op{S}^\dagger\op{a}\op{S} = \op{a}\cosh r + \op{b}^\dagger\sinh r$, it follows that taking $r\to -r$ gives $\op{S}\op{a}^\dagger\op{S}^\dagger = \op{a}\cosh r - \op{b}^\dagger\sinh r$. Substituting into the displacement operator:
\begin{align}
    \op{S}\op{D}_a(\alpha)\op{S}^\dagger 
    &= \exp\left[\alpha\op{S}\op{a}^\dagger\op{S}^\dagger 
    - \alpha^*\op{S}\op{a}\op{S}^\dagger\right] \nonumber\\
    &= \exp\Big[\alpha(\sqrt{G}\,\op{a}^\dagger 
    - \sqrt{G-1}\,\op{b}) \nonumber\\
    &\qquad - \alpha^*(\sqrt{G}\,\op{a} 
    - \sqrt{G-1}\,\op{b}^\dagger)\Big] \nonumber\\
    &= \exp\left[\sqrt{G}(\alpha\op{a}^\dagger 
    - \alpha^*\op{a})\right] \nonumber\\
    &\qquad\times\exp\left[\sqrt{G-1}(\alpha^*\op{b}^\dagger 
    - \alpha\op{b})\right] \nonumber\\
    &= \op{D}_a(\sqrt{G}\alpha)\,\op{D}_b(\sqrt{G-1}\,\alpha^*),
\end{align}
where the exponents can be separated since $\op{a}$ and $\op{b}$ act on different factors of the Hilbert space and therefore commute. The action of $\op{S}$ on the two-mode vacuum produces the two-mode squeezed vacuum (TMSV) state \citep[p.~85]{WallsMilburn2008}:
\begin{equation}
    \op{S}\ket{0,0} = \frac{1}{\sqrt{G}}\sum_{n=0}^\infty
    \left(\frac{\sqrt{G-1}}{\sqrt{G}}\right)^n\ket{n,n}.
\end{equation}
Combining these results, the total output state is:
\begin{multline}
    \op{\rho}_{out,total} =
    \op{D}_a(\sqrt{G}\alpha)\,\op{D}_b(\sqrt{G-1}\,\alpha^*) \\
    \times\,\ketbra{TMSV}{TMSV}\,
    \op{D}_b^\dagger(\sqrt{G-1}\,\alpha^*)
    \op{D}_a^\dagger(\sqrt{G}\alpha).
\end{multline}
We are interested in the state of the signal mode alone. Since we have no access to the internal mode $\op{b}$ of the amplifier (it is an internal degree of freedom that we cannot measure) we obtain the reduced state of the signal by tracing out mode $\op{b}$. For a total state $\op{\rho}_{total}$ acting on $\mathcal{H}_a \otimes \mathcal{H}_b$, the reduced state of mode $\op{a}$ is defined as:
\begin{equation}
    \op{\rho}_a = \mathrm{Tr}_b\left[\op{\rho}_{total}\right] 
    = \sum_n \bra{n}_b\op{\rho}_{total}\ket{n}_b,
\end{equation}
which sums over all states of mode $\op{b}$, leaving only the signal mode. Since $\op{D}_b$ acts only on mode $\op{b}$, it is absorbed into the trace and does not affect the signal mode. Furthermore, tracing out one mode of the TMSV state yields a thermal state \citep[p.~85]{WallsMilburn2008}:
\begin{equation}
    \mathrm{Tr}_b\left[\ketbra{TMSV}{TMSV}\right] = 
    \op{\rho}_{th}(\nbar),
\end{equation}
with mean photon number $\nbar = G - 1$. The final output density matrix is therefore:
\begin{equation}\label{eq:linear_amp_output}
    \op{\rho}_{out} = \op{D}(\sqrt{G}\,\alpha)\,
    \op{\rho}_{th}(G-1)\,\op{D}^\dagger(\sqrt{G}\,\alpha).
\end{equation}
This is a displaced thermal state: a coherent amplitude $\sqrt{G}\,\alpha$ representing the amplified signal, displaced on top of a thermal state with mean photon number $\nbar = G-1$ representing the minimum noise added by the amplification, in accordance with the Caves limit. 

This construction can be generalized to a non-vacuum internal-mode state; in particular, when
the internal mode $\op{b}$ starts in a thermal state rather than vacuum, the amplifier map is still
given by a two-mode squeezing operation on the extended ancilla state, as established generally
for phase-preserving linear amplifiers in Ref.~\cite{Caves2012}. Since the squeezing Hamiltonian
$r(\op a\op b - \op a^\dagger\op b^\dagger)$ is symmetric under exchanging $\op{a}$ and $\op{b}$, the same
calculation as above, with the roles of $\op{a}$ and $\op{b}$ interchanged, gives
\begin{equation}
    \op S\op D_b(\nu)\op S^\dagger = \op D_b(\sqrt{G}\nu)\,\op D_a(\sqrt{G-1}\,\nu^*).
\end{equation}
Writing the thermal state of mode $\op{b}$ in its Glauber--Sudarshan $P$-representation,
$\hat\rho_{\rm th}(n_{\rm th}) = \int d^2\nu\,P_{\rm th}(\nu)\ket{\nu}\bra{\nu}$ with
$P_{\rm th}(\nu) = e^{-|\nu|^2/n_{\rm th}}/(\pi n_{\rm th})$, and tracing out mode $\op{b}$ as before,
each coherent-state component $\ket{\nu}$ of the ancilla contributes an additional random displacement $\sqrt{G-1}\,\nu^*$ to the output of mode $\op{a}$, on top of the displaced thermal state $\op\rho_{\rm th}(G-1)$ already obtained for a vacuum ancilla. Averaging over the thermal mixture of $\nu$ is a convolution of two independent, circularly symmetric Gaussian $P$ functions in phase space, which simply adds their variances. The result is again a displaced thermal state, now with
\begin{equation}\label{eq:nbar_thermal_ancilla}
    \op\rho_{\rm out} = \op D(\sqrt{G}\,\alpha)\,
    \op\rho_{\rm th}\!\big[(G-1)(n_{\rm th}+1)\big]\,
    \op D^\dagger(\sqrt{G}\,\alpha),
\end{equation}
reducing to Eq.~\eqref{eq:linear_amp_output} when $n_{\rm th}=0$.

A displaced thermal state 
$\op{D}(\beta)\op{\rho}_{th}(\nbar)\op{D}^\dagger(\beta)$ 
has mean photon number $\expval{\op{n}} = \nbar + |\beta|^2$ 
and variance:
\begin{equation}
    \expval{\op{n}^2} - \expval{\op{n}}^2 = 
    \nbar^2 + \nbar + 2\nbar|\beta|^2 + |\beta|^2.
\end{equation}
Using the definition of the Mandel $Q$ parameter:

\begin{equation}\label{eq:mandelQ}
    Q = \frac{\expval{\op{n}^2} - \expval{\op{n}}^2}
    {\expval{\op{n}}} - 1,
\end{equation}we can see that the Mandel $Q$ parameter of a displaced thermal state evaluates to:
\begin{equation}\label{eq:Q_displaced_thermal}
    Q = \frac{\nbar\left(\nbar + 2|\beta|^2\right)}
    {\nbar + |\beta|^2}.
\end{equation}
For a pure coherent state ($\nbar = 0$) this gives $Q = 0$, 
and for a thermal state with no displacement ($|\beta| = 0$) 
it gives $Q = \nbar$, both as expected.

\section{Purity of the Amplified Cat State}
To find the purity of the Amplified cat state mentioned in section 3, we first calculate the Wigner function explicitly.

\subsection{Wigner Function of the Amplified Cat State}
We calculate the Wigner function of the amplified state, Eq. \eqref{eq:rhoK}, which
inherits its decomposition linearly:
\begin{equation}
    W = \tfrac{1}{2}W_{\rm self}^{+} + \tfrac{1}{2}W_{\rm self}^{-} + W_{\rm int},
\end{equation}
where $W_{\rm self}^{\pm}$ come from the coherent-state terms
$\ket{\pm i\beta}\bra{\pm i\beta}$ and $W_{\rm int}$ from the cross terms
$\ket{\pm i\beta}\bra{\mp i\beta}$, with $\beta = \eta - \sqrt{G}i\alpha$.

\textit{Coherent-state outer products.}

Using the standard definition $W_{\hat A}(x,p) = \frac{1}{\pi\hbar}\int dy\,
\bra{x-y}\hat A\ket{x+y}e^{2ipy/\hbar}$ and the position-representation
overlap $\braket{x}{\beta}$, a short Gaussian-integral calculation gives,
for $\hat A = \ket{\beta_1}\bra{\beta_2}$,
\begin{align}\label{eq:app_Wcoh}
    W_{\ket{\beta_1}\bra{\beta_2}}(x,p) &= \frac{1}{\pi\hbar}
    \exp\!\Big[-\tfrac{x^2+p^2}{\hbar}
    + \sqrt{\tfrac{2}{\hbar}}(\beta_1+\beta_2^*)x \notag\\
    &\quad + i\sqrt{\tfrac{2}{\hbar}}(\beta_2^*-\beta_1)p
    - \beta_2^*\beta_1 - \tfrac{|\beta_1|^2+|\beta_2|^2}{2}\Big].
\end{align}
Rescaling to phase-space coordinates in which $\ket{\beta}$ is centred at
$(\mathrm{Re}\,\beta,\mathrm{Im}\,\beta)$ (i.e. $x\to x\sqrt{2\hbar}$,
$p\to p\sqrt{2\hbar}$) gives the diagonal and off-diagonal forms
\begin{align}
    W_{\ket{\beta}\bra{\beta}}(x,p) &= \frac{2}{\pi}
    \exp\!\big[-2(x-\mathrm{Re}\,\beta)^2 - 2(p-\mathrm{Im}\,\beta)^2\big],
    \label{eq:app_Wdiag}\\
    W_{\ket{\beta_1}\bra{\beta_2}}(x,p) &= \frac{2}{\pi}
    \exp\!\Big[-2(x^2+p^2) + 2(\beta_1+\beta_2^*)x \notag\\
    &\qquad + 2i(\beta_2^*-\beta_1)p - \beta_2^*\beta_1
    - \tfrac{|\beta_1|^2+|\beta_2|^2}{2}\Big].
    \label{eq:app_Woffdiag}
\end{align}

\textit{Self-terms.}
Applying Eq.~\eqref{eq:app_Wdiag} to $\ket{\mp i\beta}\bra{\mp i\beta}$ and
averaging over the thermal weight $e^{-|\eta|^2/\bar n}/(\pi\bar n)$ (two
decoupled Gaussian integrals in $\eta_r,\eta_i$) gives
\begin{equation}\label{eq:app_Wself}
    W_{\rm self}^{\pm}(x,p) = \frac{2}{\pi(1+2\bar n)}
    \exp\!\left(-\frac{2\big[(x \pm \sqrt{G}\alpha)^2 + p^2\big]}{1+2\bar n}\right),
\end{equation}
two Gaussian blobs centred at $(x,p)=(\mp\sqrt G\alpha,0)$.

\textit{Interference term.}
Writing $W^{+-}\equiv W_{\ket{i\beta}\bra{-i\beta}}$, the cross-term
contribution is $W_{\rm int} = \mathrm{Re}[-i\,W^{+-\,(\rm avg)}]$. Since
$\beta_2^*\beta_1 = -|\beta|^2$ for this pair, the overlap prefactor in
Eq.~\eqref{eq:app_Woffdiag} vanishes identically, leaving
\begin{equation}
    W^{+-}(x,p) = \frac{2}{\pi}e^{-2(x^2+p^2)}
    \exp\!\big[4i(\eta_r x + (\eta_i-\sqrt{G}\alpha)p)\big].
\end{equation}
Thermally averaging over $\eta_r,\eta_i$ (again two decoupled Gaussian
integrals, each contributing a factor $\sqrt{\pi\bar n}$) gives
$W^{+-\,(\rm avg)}(x,p) = \frac{2}{\pi}e^{E_R+iE_I}$ with
$E_R = -2(1+2\bar n)(x^2+p^2)$ and $E_I = -4\sqrt{G}\alpha p$. Hence
\begin{equation}\label{eq:app_Wint}
    W_{\rm int}(x,p) = -\frac{2}{\pi}
    \exp\!\big[-2(1+2\bar n)(x^2+p^2)\big]\sin\!\big(4\sqrt{G}\alpha\,p\big),
\end{equation}
with $\bar n = G-1$. The full Wigner function of the amplified cat state is
therefore
\begin{equation}\label{eq:app_Wfull}
    W(x,p) = \tfrac{1}{2}W_{\rm self}^{+}(x,p) + \tfrac{1}{2}W_{\rm self}^{-}(x,p)
    + W_{\rm int}(x,p),
\end{equation}
with $W_{\rm self}^\pm$ given by Eq.~\eqref{eq:app_Wself} and $W_{\rm int}$
by Eq.~\eqref{eq:app_Wint}.

\subsection{Purity of the Amplified Cat State}
\label{app:purity}

The purity follows from the Wigner function via
$\mathrm{Tr}(\hat\rho^2) = \pi\int W^2\,dx\,dp$. Substituting
Eq.~\eqref{eq:app_Wfull} and expanding the square,
\begin{align}
    \int W^2\,dx\,dp &= \tfrac{1}{4}\!\int (W_{\rm self}^{+})^2
    + \tfrac{1}{4}\!\int (W_{\rm self}^{-})^2 \notag\\
    &\quad + \tfrac{1}{2}\!\int W_{\rm self}^{+}W_{\rm self}^{-}
    + \int W_{\rm int}^2 \notag\\
    &\quad + \int W_{\rm int}\big(\tfrac{1}{2}W_{\rm self}^{+}+\tfrac{1}{2}W_{\rm self}^{-}\big).
\end{align}
In the large-$|\alpha|$ limit the two self-term Gaussians are well
separated, so the cross term $\int W_{\rm self}^{+}W_{\rm self}^{-}$
vanishes; the interference term $W_{\rm int}$ also oscillates rapidly
in $p$ while the self-terms are smooth, so the remaining cross terms
vanish as well, leaving
\begin{equation}\label{eq:app_purity_split}
    \int W^2\,dx\,dp \approx \tfrac{1}{4}\!\int (W_{\rm self}^{+})^2
    + \tfrac{1}{4}\!\int (W_{\rm self}^{-})^2
    + \int W_{\rm int}^2 .
\end{equation}

\textit{Self-term contribution.}
Each self-term is a Gaussian of width $\tfrac{1+2\bar n}{4}$ and
prefactor $\tfrac{2}{\pi(1+2\bar n)}$, giving
$\int (W_{\rm self}^{\pm})^2\,dx\,dp = \tfrac{1}{\pi(1+2\bar n)}$.
The combined self-term contribution to the purity is therefore
\begin{equation}
    \pi\cdot\tfrac{1}{2}\cdot\tfrac{1}{\pi(1+2\bar n)}
    = \frac{1}{2(1+2\bar n)} = \frac{1}{2(2(G-1)(n_{th} + 1))}.
\end{equation}

\textit{Interference contribution.}
Squaring Eq.~\eqref{eq:app_Wint} and using $\sin^2\theta =
\tfrac{1}{2}(1-\cos 2\theta)$, the $x$ and $p$ integrals are standard
Gaussians:
\begin{align}
    \int W_{\rm int}^2\,dx\,dp
    &= \frac{4}{\pi^2}\int e^{-4(1+2\bar n)x^2}\,dx \notag\\
    &\quad \times \int e^{-4(1+2\bar n)p^2}
    \sin^2\!\big(4\sqrt{G}\alpha\,p\big)\,dp \notag\\
    &= \frac{1}{2\pi(1+2\bar n)}\Big(1 - e^{-4G\alpha^2/(1+2\bar n)}\Big).
\end{align}
The corresponding purity contribution is
\begin{multline}
    \pi\cdot\frac{1}{2\pi(1+2\bar n)}\Big(1-e^{-4G\alpha^2/(1+2\bar n)}\Big) \\
    = \frac{1}{2(1+2\bar n)}\Big(1-e^{-4G\alpha^2/(1+2\bar n)}\Big),
\end{multline}
which approaches $\tfrac{1}{2(2(G-1)(n_{th}+1))}$ in the large-$|\alpha|$ limit,
where the exponential vanishes.

\textit{Result.}
Adding the two contributions,
\begin{multline}
    \mathrm{Tr}(\hat\rho^2) = \frac{1}{2\big(2(G-1)(n_{th}+1)\big)} \\
    + \frac{1}{2\big(2(G-1)(n_{th}+1)\big)}
    = \frac{1}{2(G-1)(n_{th}+1)},
\end{multline}
exactly saturating the Caves-limit purity bound of Sec.~III. Any
larger interference amplitude at fixed $\bar n$ would increase
$\int W_{\rm int}^2$ beyond what this bound allows.

\section{Kerr-Cat Splitting Identity}
\label{app:kerr-splitting}

Expanding a coherent state in the Fock basis,
$\ket{\beta} = e^{-|\beta|^2/2}\sum_{n=0}^\infty
\frac{\beta^n}{\sqrt{n!}}\ket{n}$, and applying the Kerr unitary
$\op{U}_{\rm Kerr}(\kappa t)$ of Eq.~\eqref{eq:kerr_unitary}, which is diagonal in this basis, gives
\begin{equation}
    \op{U}_{\rm Kerr}(\kappa t)\ket{\beta} = e^{-|\beta|^2/2}
    \sum_{n=0}^\infty \frac{\beta^n}{\sqrt{n!}}
    e^{-i\frac{\kappa t}{2}n(n-1)}\ket{n}.
\end{equation}
At $\kappa t = \pi$, the phase factor reduces to
$e^{-i\frac{\pi}{2}n(n-1)} = (-1)^{n(n-1)/2}$, since $n(n-1)$ is always even. This sequence has period 4 in $n$
($+1,+1,-1,-1,\ldots$) and can therefore be decomposed as
$a + bi^n + c(-1)^n + d(-i)^n$; solving for the coefficients gives
$a=c=0$ and
\begin{equation}\label{eq:app_phase_decomp}
    e^{-i\frac{\pi}{2}n(n-1)} = \frac{e^{-i\pi/4}}{\sqrt2}i^n
    + \frac{e^{i\pi/4}}{\sqrt2}(-i)^n.
\end{equation}
Substituting Eq.~\eqref{eq:app_phase_decomp} into the evolved state and resumming each half as a coherent state in the Fock basis yields the cat-splitting identity used in Eq.~\eqref{eq:catsplit}:
\begin{equation}
    \op{U}_K\ket{\beta} = \frac{e^{-i\pi/4}}{\sqrt2}\ket{i\beta}
    + \frac{e^{i\pi/4}}{\sqrt2}\ket{-i\beta},
\end{equation}
a two-component cat state with lobes centred at $\pm i\beta$
\cite{YurkeStoler1986}. More generally, a coherent state propagating through a Kerr medium evolves into an $N$-kitten state at
$\kappa t = 2\pi M/N$ for coprime integers $M,N$
\cite{Tara1993,Tanas2003},
\begin{align}
    \ket{\psi_{M,N}} &= \sum_{k=0}^{N-1} f_k
    \ket{\beta e^{ik\pi/N}}, \\
    f_k &= \frac{1}{2N}\sum_{n=0}^{2N-1}
    e^{-i\frac{\pi}{N}[nk-Mn(n-1)]}.
\end{align}
\cite{Propp2023}. The optimality proof of Sec.~III applies to all such states, since each is unitarily generated from a coherent state.

\end{document}